# Uncovering electron scattering mechanisms in NiFeCoCrMn derived concentrated solid solution and high entropy alloys

Sai Mu[1], German. D. Samolyuk[1], Sebastian. Wimmer[2], Maria C. Troparevsky[1], Suffian N. Khan[1], Sergiy Mankovsky[2], Hubert Ebert[2], George M. Stocks[1*]

[1] Materials Science & Technology Division, Oak Ridge National Laboratory, Oak Ridge, TN 37831, USA

[2] Department of Chemistry, Ludwig-Maximilians-Universität, D-81377 München, Germany

AUTHOR INFORMATION

Corresponding Author

**Sai Mu,** Material Science and Technology Division, Oak Ridge National Lab, P. O. Box 2008, Oak Ridge, TN, 38831-6114 (tel: 402-890-9490) (sai.mu1986321@gmail.com)

**G. M. Stocks**, Material Science and Technology Division, Oak Ridge National Lab, P. O. Box 2008, Oak Ridge, TN, 38831-6114 (stocksgm@ornl.gov)

# Abstract

Whilst it has long been known that disorder profoundly affects transport properties, recent measurements on a series of solid solution *3d*-transition metal alloys reveal two orders of magnitude variations in the residual resistivity. Using *ab-initio* methods, we demonstrate that, while the carrier density of all alloys is as high as in normal metals, the electron mean-free-path can vary from ~10 Å (strong scattering limit) to ~$10^3$ Å (weak scattering limit). Here, we delineate the underlying electron scattering mechanisms responsible for this disparate behavior. While site-diagonal, spin dependent, potential scattering is always dominant, for alloys containing only Fe, Co, and Ni the majority spin channel experiences negligible disorder scattering, thereby providing a short circuit, while for Cr/Mn containing alloys both spin channels experience strong disorder scattering due to an electron filling effect. Somewhat surprisingly, other scattering mechanisms – including displacement, or size effect, scattering which has been shown to strongly correlate with such diverse properties as yield strength – are found to be relatively weak in most cases.

# Introduction

Electrical resistivity is one of the most fundamental properties of materials. At the coarsest level, it distinguishes between metals, semi-conductors and insulators. As such, it provides a window into the properties of the electron glue responsible for cohesion. In metals and alloys the electrical resistivity is directly related to the mean free path, $\lambda_e[\varepsilon_F]$, (alternatively the lifetime, $\tau_e[\varepsilon_F]$) of electrons at the Fermi energy. In a pure crystalline metal at absolute zero of temperature (T = 0 K), Bloch states are eigenstates of the system, $\lambda_e[\varepsilon_F]$ and $\tau_e[\varepsilon_F]$ are infinite, and the resistivity vanishes. In disordered solid solution alloys, the chemical disorder that results from the random distribution of the alloying elements on the underlying crystalline lattice induces electron scattering and finite $\lambda_e[\varepsilon_F]$ and $\tau_e[\varepsilon_F]$ even at absolute zero. As a result, the T = 0 K resistivity, or residual resistivity $\rho_0$, is finite and its precise value provides a direct measure of the disorder induced changes in the underlying electronic structure.

In a general N-component solid solution alloy the chemical disorder, as measured by the ideal entropy of mixing, is maximal at equiatomic composition and increases with the number of components. Equiatomic High Entropy Alloys (HEA), are exemplars of such maximally disordered alloys in that they are comprised of N≥5 components yet unexpectedly form highly stable, single-phase, disordered solid-solutions on a simple crystal lattice. The first single-phase HEA, NiFeCoCrMn, was synthesized by Cantor et al.[1,2] in 2004. Since then HEAs have become a subject of intense scientific and technological interest[3–6]. In 2014, Wu et al.[7], showed that alloying the elements of Cantor's alloy (supplemented with Pd) yields a series of 2-, 3-, 4-component equiatomic *fcc* solid solutions: NiPd, NiCo, NiFe, NiFeCo, NiCoCr, NiCoMn, NiCrCoMn, NiFeCoMn and NiFeCoCr. This set of alloys combined with NiFeCoCrMn and NiFeCoCrPd (here collectively referred to as Cantor-Wu alloys), constitute a rich playground for comprehensive studies of the role of maximal disorder on the properties of multi-component alloys by controlling both the number (increasing configurational entropy) and types (chemical specificity) of alloying elements[4,5,8].

Of interest here are the results of recent residual resistivity measurements[5,8] of a subset of Cantor-Wu alloys that show, rather than increasing monotonically with increasing numbers of components, values of $\rho_0$ break into two subgroups of low ($\rho_0$ <10 µΩ•cm) and high ($\rho_0$ >75 µΩ•cm) resistivity alloys. In addition, two entropically identical alloys, NiCoFe ($\rho_0$ = 1.7µΩ•cm) and NiCoCr ($\rho_0$ =92.7 µΩ•cm), fall into different resistivity groupings. Remarkably, the least and most resistive alloys differ by almost two orders of magnitude, $\rho_0$(NiCo)=1.3µΩ•cm; $\rho_0$(NiFeCoCrPd)=124.8µΩ•cm. Interestingly, the low resistivity group have $\rho_0$ values typical of dilute weak scattering alloys in which there are clearly defined host (solvent) and impurity (solute) elements. In such alloys, $\rho_0$ arises from the scattering of a low Fermi energy DOS of nearly-free-electron *sp*-states with large $\lambda_e[\varepsilon_F]$ and $\rho_0$ generally obeys both Nordheim's relation ($\rho_0 \propto c((1-c)$; where *c* is impurity concentration)[9] and Linde's "law" ($\rho_0 \propto (\Delta Z)^2$; where $\Delta Z$ is the valence difference between host and impurity atoms)[10]. (see Ref.[11] for a discussion) This, despite the fact that, in equiatomic alloys, the concept of host and impurity elements is lost and the Fermi energy falls in the high density of state (DOS) *d*-bands[5]. At the other extreme, high-$\rho_0$ NiFeCoCrPd is close to the Mott-Ioffe-Regel (MIR) limit[12,13], which is characterized by a $\lambda_e[\varepsilon_F]$ value comparable to the lattice spacing[12,14,15]. Combined, these observations suggest that, although the Cantor-Wu alloys are highly crystalline and have uniformly high Fermi energy carrier densities, $\lambda_e[\varepsilon_F]$ can be controlled, from ~10 Å to ~$10^3$ Å, by the specifics of the number and types of alloying elements. Furthermore, distinct from many other metallic materials with high resistivity, the disorder-induced short $\lambda_e[\varepsilon_F]$ of highly resistive Cantor-Wu alloys does not require strong electron correlation as in incoherent metals[16–18], large atomic displacements associated with very high temperatures, or complete loss of translational symmetry as in quasicrystals[19]. As such the Cantor-Wu alloys provide a unique opportunity for uncovering the underlying scattering mechanisms that give such disparate and non-monotonic behavior in *3d*-transition metal alloys that form on well-defined, in this case *fcc*, crystalline lattices.

Here, we report the first calculations of the residual resistivity of the full set of Cantor-Wu alloys using state-of-the-art *ab initio* transport theory for disordered alloys. Consistent with experiment, we find that the calculated $\rho_0$ break into high-$\rho_0$ (alloys involving Mn/Cr elements) and low-$\rho_0$ (the others) sets. We show that it is the magnitude of the spin-dependent site-diagonal potential scattering that makes the dominant contribution to $\rho_0$ and gives rise to this remarkable difference in $\rho_0$ between the two sets. We explicitly evaluate effects of disorder that go beyond those captured by conventional CPA – local lattice displacements, the distribution of both the magnitude and orientation of the local magnetic moments. Surprisingly, we find that the scattering from local lattice distortions as well as the site-to-site variations in local moment magnitude and orientation are relatively weak in most alloys. This despite the fact that one of these, lattice displacements, has been shown to strongly correlate with such a seemingly unrelated property as yield strength.

**Results and Discussions**

In solid solution alloys, all electron scattering ultimately results from the disorder-induced site-to-site potential fluctuations. However, to understand the fundamentals of the scattering mechanisms, it is useful to divide the total scattering according to a number of distinguishable sub-mechanisms. Single-site electron scattering can be thought of as resulting from the site-to-site variation ($\delta$) in the local potential due to the random distribution of elements. In the presence of magnetism, conduction electrons experience an additional inhomogeneous exchange field ($\Delta_{Exch}$), which further increases the site-disorder and is different in separate spin channels. In the following, we shall refer to *site-diagonal disorder* as being the combined effects of [$\delta$, $\Delta_{Exch}$]. The single-site picture is further modified by including the effects of *displacement scattering* caused by relaxation of the atoms away from their ideal lattice sites due to the fact that every atom is surrounded by a different configuration of other atoms. Moreover, additional magnetic scattering can arise from fluctuations about the species-dependent single-site average in both the size of the local moments and how they couple amongst themselves – ferromagnetic, antiferromagnetic, mixed ferro/antiferro, non-collinear, non-coplanar …

***Site-diagonal disorder.*** – Using the conventional spin-polarized KKR-CPA method, we explore the effect of site-diagonal disorder, *i.e.* [$\delta_\sigma$, $\Delta_{Exch}$], on the electronic structure and $\rho_0$ in Cantor-Wu alloys. Figure 1 compares the calculated $\rho_0$ with the measured values[5,8]. From the figure, there are three clear conclusions. Firstly, consistent with the experiments, the calculated values of $\rho_0$ separate into two groups: low-$\rho_0$ alloys (NiPd, NiCo, NiFe, NiFeCo), having $\rho_0 < 10$ μΩ•cm and high-$\rho_0$ alloys (the others). This finding is independent of the particular exchange correlation functional used. Secondly, while the calculated value of $\rho_0$ including only site-diagonal disorder, underestimates $\rho_0$, the contribution from site-diagonal disorder is dominant across all Cantor-Wu alloys. Thirdly, the magnitude of $\rho_0$ correlates with the types of alloying elements. In particular, for alloys containing only the Ni, Fe, Co, that have nearly-filled *3d*-bands, $\rho_0$ is low. While for alloys containing both Ni, Fe, Co, and Cr, Mn, whose *d* bands are approximately half-filled, $\rho_0$ is large. Notably, the latter set of alloys are also characterized by mixed exchange coupling between the local moments of Ni, Fe, Co (ferromagnetic) and Cr, Mn (antiferromagnetic) while the former exhibit only ferromagnetic coupling.

The underlying reason for the breakdown into two distinct resistivity groups can be understood in terms of disorder smearing of the Fermi surface. Figure 2 (a) shows the spin-resolved Fermi surfaces of four selected Cantor-Wu alloys – two each from the low-$\rho_0$ and high-$\rho_0$ group. While the minority-spin Fermi surfaces exhibit large disorder smearing for all of the alloys, the majority-spin channels are very different in the two classes. In particular, the majority-spin Fermi surfaces for NiCo and NiFeCo remain very sharp which corresponds to a long $\lambda_e[\varepsilon_F]$. As a result, the majority spin channel acts as a short circuit for electron conduction resulting in an overall low resistivity. On the contrary, the majority-spin Fermi surfaces of NiFeCoCr and NiFeCoCrMn alloys are washed out with the consequence that the $\lambda_e[\varepsilon_F]$ in both spin channels is very short and thus $\rho_0$ is high. In the absence of a direct calculation of the residual resistivities, it has been previously noted that the transport properties of the Cantor-Wu alloys qualitatively reflect the large differences in disorder smearing of the Fermi energy Bloch spectral functions[5,8], that are driven by differences in magnetic (FM versus mixed FM/AFM) coupling – an conclusion that turns out to be inadequate and even misleading. Notably, NiCoCr also has a very smeared Femi energy Bloch spectral function, and correspondingly high $\rho_0$, despite being robustly nonmagnetic[20].

Figure 2 (b) shows a cartoon of the underlying spin-resolved partial DOS of the alloying elements Ni, Co, Fe, Cr that illustrates why the spin-resolved Fermi surface smearing is so different in the two alloy groups. Within the KKR-CPA, the strength of the disorder scattering can be characterized by the ratio of two important energy scales: the energy separation ($\delta$) between the band centers of different species a.k.a "band center mismatch" and the overall band width (W)[21]. In transition metals, the most relevant band center is simply the *d*-scattering resonance ($\varepsilon_d$) of the single-site potential, while W encapsulates the spread of the *d*-bands due to hybridization. These energy scales are illustrated in Fig. 2 (b). If $\delta/W \ll 1$, the disorder scattering is weak, and the electron bands are well-defined. However, if $\delta/W \sim 1$, disorder scattering is strong, leading to large disorder broadening (smearing) of the energy bands. For magnetic alloys, the electrons propagate and are scattered in two separate and independent spin channels[22] – neglecting the spin-mixing contribution. The spin-mixing arising from spin-orbit coupling however affects $\rho_0$ and brings about the anisotropic magnetoresistance as shown by Banhart *et al.*[23]. As a result, the above argument applies to each spin channel independently, distinguished by subscript $\sigma = \uparrow$ and $\downarrow$ for spin-up and spin-down.

For alloys containing only Fe, Co and Ni, the majority-spin *3d* band centers are aligned due to minimization of the kinetic energy. As a result, $\delta_\uparrow$ between all atom pairs is small and thus $\delta_\uparrow/W_\uparrow$ is always in the weak scattering regime. Because, different local moments form on different species and they couple ferromagnetically, the additional exchange splitting ($\Delta_{Exch}$) which is proportional to the size of the local moment – with proportionality constant ~ 1eV/$\mu_B$, leads to a large band center mismatch ($\delta_\downarrow$) in the minority-spin channel and consequent large disorder scattering ($\delta_\downarrow/W_\downarrow \sim 1$) (see Fig. 2(b)). A similar argument has been applied in $Ni_{35}Fe_{65}$[24], and NiCo[25]. As a result, while the majority-spin Fermi surface is well-defined and $\lambda_e[\varepsilon_F]$ is long, minority-spin channel electron transport is "blocked" by the strong disorder smearing of the Fermi surface. On the other hand, when alloying with lower band filling Cr, the band center in both channels is shifted towards the Fermi energy in order to realize charge neutrality. As a consequence, $\delta$ is large in both spin channels (large disorder scattering), thereby washing out the Fermi surface. In addition, the moments on Cr can couple either ferromagnetically or antiferromagnetically, further modifying $\delta$. However, this does not substantially diminish $\delta$, and thus $\delta/W$ remains in the strong scattering regime. Similar arguments also apply to Mn. It is worth noting that $\rho_0$ is high only if strong disorder scattering is present at the Fermi energy. For example, strong disorder scattering in NiPd notwithstanding, $\rho_0$ is low ($\rho_0$= 2.19 $\mu\Omega\bullet$cm) because the Fermi surface is mainly of *sp* character and large disorder scattering of *d* electrons does not pollute the Fermi surface in either spin channel. (see the Fermi surface in the Supplementary Information, Section 5(D), Fig. S7). Whilst, these kinds of arguments are well known[24,26–28], the way they operate in this class of alloys is particularly startling.

As noted previously, while the calculated $\rho_0$ of NiCo, NiFeCo, NiFeCoCr agree quantitatively with the experiments[5,8] (see Fig. 1), $\rho_0$ is still underestimated by a substantial fraction, particularly in NiFe, NiPd and also in some high-$\rho_0$ alloys such as NiCoCr, NiFeCoCrMn and NiFeCoCrPd. To shed light on this underestimation of $\rho_0$, we explore the effects of other scattering mechanisms – displacement scattering, magnetic scattering – beyond the single-site approximation. The supercell method is employed to explore the effect of site-to-site atomic displacement and the complex magnetic effect. (see the Method section) Figure 3 (a) illustrates the supercell of NiFeCoCrMn alloy with different species randomly distributed. The direction and length of the black arrows on each site indicate the orientation and magnitude of each local moment, respectively.

***Displacement scattering***. – Recently, atomic displacements have been shown to correlate with yield strength[29], suggesting the root mean square (*rms*) atomic displacements as a descriptor of the mechanical properties of Cantor-Wu alloys. Given that all materials properties ultimately originate from the electronic structure, studying the effect of displacement scattering on the electronic transport provides a window into their importance as a scattering mechanism – albeit only at the Fermi energy.

Based on fully relaxed supercell calculations, the statistics of the magnitude of the atomic displacements – resolved by species – are shown in Fig. 3 (c) for selected Cantor-Wu alloys (see the Supplementary Information, Section 3, Fig. S2 for other alloys). As seen in the NiFeCo ternary alloy, atomic displacements (Δu) on all species are small with $\Delta u(Ni) < \Delta u(Co) < \Delta u(Fe)$. This is consistent with the elements having similar atomic size and electronegativities. On the other hand, for alloys involving Cr and/or Mn, the atomic displacements on Cr and Mn are much larger. Again, consistent with expected larger size mismatch and charge transfer effects. As such, species-dependent displacements become more pronounced towards the left side of the *3d*-transition metal elements in the periodic table. Moreover, if alloying with Pd –as in NiPd and NiFeCoCrPd alloys, atomic displacements on all species are large due to size-mismatch between *3d* and *4d* elements (see Review article[30] for the size-mismatch effect). Notably, the statistics of the angular dependence of the displacements appears to be random (see Fig. 3 (b) for an illustration in NiFe alloy).

Assuming that the site-to-site variation of both magnitude and the orientation of the atomic displacements are uncorrelated, their effects on the electronic structure and $\rho_0$ can be assessed by the alloy analogy model (AAM)[31]. The results are shown in Figure 4 (a). In most alloys the enhancement of $\rho_0$ due to local displacements is rather small because the atomic displacements (of most alloys) are only a small fraction of the interatomic spacing. The exceptions are NiPd and NiFeCoCrPd alloys whose atomic displacements are large for all component species (see Fig. 3 (c)). For NiPd, the resulting displacement-enhanced resistivity is in good agreement with experiment. While for NiFeCoCrPd the inclusion of displacement scattering increases $\rho_0$ by ~12%, the actual $\rho_0$ is still underestimated. Therefore, the overall effect of displacement scattering in most alloys is small. Thus, the reasons for the general underestimation of $\rho_0$ by single-site theory alone must be sought elsewhere. Furthermore, this finding makes the strong correlation between the *rms* displacements and yield strength all the more interesting; perhaps, suggesting the existence of a more fundamental descriptor, rooted in the (common) underlying electronic structure.

*Magnetism beyond the single-site approximation*. – Unlike in the KKR-CPA treatment of magnetism which deals with species-dependent single-site averaged magnetic moment, in the real alloys, the local moments of each species take on a distribution of values and can possibly point along arbitrary directions. In Fig. 3 (d) we show that species-resolved local moment distributions for selected Cantor-Wu alloys, obtained using the supercell method, with moments constrained to be co-linear. For the local moment distribution in other alloys, see Supplementary Information, Section 4(A), Fig. S3. It is noteworthy that, although supercell calculations yield a distribution of local moments, the species-dependent averaged local moments turn out to be consistent with those obtained from KKR-CPA. (see the Supplementary Information, Section 4(B), Tab. S2)

For alloys considered here, several features of the local moment patterns can be found. Firstly, the local moments in low-$\rho_0$ alloys are ferromagnetically coupled and display only a small variation in the size of the moments. In high-$\rho_0$ alloys with Cr, the magnitudes of Cr moments vary widely from negative (antiferromagnetically aligned) to positive (ferromagnetically aligned). In contrast to Cr, the magnitudes of Mn moments fall into two well-defined groups, large positive and large negative. This suggests that large on-site Hund's exchange promotes the formation of local moments on Mn, while the interatomic exchange interaction between Mn atoms is antiferromagnetic. The antiferromagnetic coupling associated with Cr and Mn can be attributed to the approximately half-filled *d* bands[32–34]. Clearly, the complicated magnetic configurations just described have the potential to induce significant additional electron scattering beyond that included in the KKR-CPA[22,35–37].

Using NiFe as an example, where $\rho_0$ is underestimated ($\rho^{Calc}_0$ =3.3 versus $\rho^{Exp}_0$ =10.3 μΩ•cm), we first evaluate the effect of having a distribution of the local moment sizes. Based on supercell calculations, we find a Fe moment distribution that has a 0.2 $\mu_B$ broadening. To mimic these moment fluctuations using CPA, we discretize the Fe moment distribution into three types, having moments $m_0$ and $m_0 \pm 0.1 \mu_B$ (where $m_0$ is the average Fe moment), through scaling of the spin part of the exchange-correlation functional. Unsurprisingly, $\rho_0$ only increases by 0.3 μΩ•cm, indicating the $\rho_0$ is insensitive to such longitudinal spin fluctuations.

Taking NiCoMn and NiFeCoCrMn alloys as examples, it is noted that supercell calculations find two well-separated Mn moment distributions with opposite orientations and roughly similar amounts of Mn↑ (~ 65%-35%) and Mn↓ (~ 35%-65%) moments as the ground state. Due to the very localized nature of the Mn local moments such states can also be described within the KKR-CPA approach in a manner analogous to disordered local moment state (DLM)[38,39] previously used to describe paramagnetic Fe. Rather than investigate all possible ratios of Mn↑ and Mn↓, we focus the system having equal concentrations as an approximated representative of the state having maximal "Mn-moment" disorder – we denote this state as DLM-Mn. Similarly, another collinear magnetic state with 100% Mn↓, and one fully DLM state (DLM on all species) are also found to be stable solutions. Table 1 lists the CPA total energies for the three states. As expected, the DLM-Mn state of NiCoMn is the ground state. Moreover, the DLM-Mn state of NiCoMn gives ±2.2 $\mu_B$ for the local moment of Mn↑ and Mn↓, which is consistent with the averaged Mn↑ and Mn↓ moments (~ ±2.3 $\mu_B$) obtained from the supercell calculation. However, the Mn moment within the AFM state is only -0.7 $\mu_B$, further casting suspicion on 100% Mn↓ as representative of Mn containing Cantor-Wu alloys. Further justification of the efficacy of the DLM-Mn state can be obtained by comparing the species-resolved DOS with the supercell and studying stability of the Heisenberg interactions[40]. For $\rho_0$, we find its value depends sensitively on the assumed magnetic state – $\rho_0$ is lowest in the AFM state and increases by ~50% in the DLM-Mn ground state. Unfortunately, the experimental value has not yet been measured. The reason for the different $\rho_0$ behavior can be easily traced to the underlying electronic structure: the AFM state exhibits a relatively sharp Fermi surface in the minority-spin channel while the Fermi surface in both spin channels smears out for the DLM-Mn state (see the Supplementary Information, Section 5(B), Fig. S6). In contrast to NiCoMn, it turns out the AFM and DLM-Mn states in NiFeCoCrMn alloy are not only close in energy but their $\rho_0$ are insensitive to which state is considered because the electron scattering by magnetic-driven disorder is already almost saturated.

So far, $\rho_0$ has been calculated assuming collinear spin configurations. However, spin noncollinearity is also possible, particularly in Mn- and/or Cr- containing alloys due to the geometric frustration of antiferromagnetism on a triangular lattice (as the (111) plane of the *fcc* lattice)[41–43], oscillating exchange interactions as a function of distance[28,44], and spin-orbit interaction. For example, NiFeCoCrMn alloy is found to have a spin glass state both experimentally[45] and theoretically[46].

In principle, the spin noncollinearity can be dealt with straightforwardly from the spin noncollinear calculations based on the supercell method. However, such calculations for supercell size sufficient to gain good statistics of distributions of spin orientations are extremely demanding and remain a research project in their own right. Such calculations are made particularly difficult by the need not only treat the spin noncollinearity but also include the spin-orbit interactions – particularly for Pd-containing Cantor-Wu alloys, making them beyond the scope of this paper.

Without such a sophisticated evaluation of the spin noncollinearity at zero temperature (probability distribution of the spin orientations for each species), here we calculate the *maximum* contribution to the residual resistivity ($\rho_0$) that can arise from spin disorder. To assess this, we again employ the AAM by using a discrete set of random and uncorrelated spin moments that are distributed uniformly in space, where the magnitude of the species-dependent local moments are obtained from CPA ground state[31]. Notably, spin noncollinearity in low-$\rho_0$ alloys is negligibly small, as verified by fully relativistic supercell calculations. Therefore, we only explore the effect of full spin disorder in the high-$\rho_0$ alloys. The resulting resistivities, which can be viewed as the maximum effect of spin disorder on $\rho_0$, are shown in Fig. 4 (b). A sizable $\rho_0$ enhancement, as large as 10 μΩ•cm, is observed in NiFeCoCr, NiFeCoMn, and NiFeCoCrPd. Therefore, the full spin disorder produces a modest increase of $\rho_0$ in high-$\rho_0$ alloys.

Notwithstanding the overall improved agreement with the measurement resulting from inclusion of the additional scattering mechanisms discussed above, the remaining moderate underestimation of $\rho_0$ in some alloys (NiFe, NiCoCr, NiFeCoCrMn, NiFeCoCrPd) requires consideration of other possible theoretical shortcomings. The three most obvious being: going beyond the single-site treatment of disorder; inclusion of any possible (but currently mostly unknown except for NiFe[47]) short range order (SRO); consideration of additional electron correlation effects, beyond LDA. For binary alloys where the first two effects have been considered their impact on $\rho_0$ has been found to be small[48,49] (with specific exceptions, e.g., in K-state alloys[50]). However, it is not clear whether the impacts of the first two effects in the Cantor-Wu alloys are also small. Specifically, the specific heat measurements of Cr-containing Cantor-Wu alloys[51] show a K-state transition at 800-1000 K, which is usually attributed to the order-disorder transition[52–54]. Although the effects of SRO on $\rho_0$ are clearly worthy investigating, accounting for them requires treatments of multisite scattering processes that go beyond the single-site approximation – e.g. nonlocal CPA (see Review article[55] and references therein) – and are beyond current capabilities. In principle the investigation of additional Coulomb correlation effects on $\rho_0$ could be addressed by a combination of KKR-CPA and Dynamical Mean Field Theory[56], however, it is beyond the scope of the present work.

In conclusion, we have demonstrated that the abnormal and disparate electronic transport in Cantor-Wu alloys at zero temperature is dominated by electron scattering arising from site-to-site potential disorder. In particular, it is found that Cr and/or Mn produce strong disorder scattering as a result of the proximity of the *d*-scattering resonance to the Fermi energy which results from their reduced band filling. Additionally, other electron scattering mechanisms are explored explicitly and shown to be small in most alloys with the exception of: NiPd and NiFeCoCrPd, where the effect of displacement scattering is large; NiCoMn, where the DLM-Mn ground state significantly raises $\rho_0$; and NiFeCoCrMn/NiFeCoCrPd, where the effect of spin-noncollinearity is large. A profound understanding of the electronic transport in disordered alloys not only provides insights on the Fermi surface, but also on the overall effects of disorder throughout the occupied *d*-bands. As such it sheds light on the bonding mechanisms responsible for many of the exotic properties of HEAs, such as mechanical[4,57], and radiation response[5] properties. Here it is notable that while the effect of displacement scattering on $\rho_0$ is small, the highly unusual solid solution strength in several of Cantor-Wu alloys has been correlated with the magnitude of displacement fluctuations[29]. Resolving this apparent dichotomy is clearly a challenge worthy further investigation.

**Method**

Our calculations employ the *ab-initio* spin-polarized fully relativistic Korringa-Kohn-Rostoker Green's function method[58,59], combined with the Coherent Potential Approximation[60] (hereafter referred to as KKR-CPA) – as implemented in the Munich SPR-KKR package[61,62] – to calculate the effect of disorder on the electronic structure and residual resistivity. The CPA method is a self-consistent theory of the effects of substitutional disorder on the configurationally averaged single-site properties of alloys. As such it includes, the effect of [$\delta$, $\Delta_{Exch}$] on electronic band structure and quasiparticle lifetime on the same *ab-initio* footing. The conductivity tensor is calculated by using the linear response Kubo-Greenwood formula[63,64] for the configurational averaged state, described by the CPA[65]:

$$\sigma_{\mu\nu} = \frac{\hbar}{\pi N\Omega} Tr\{< j_\alpha^\mu ImG^+(E_F) j_\beta^\nu ImG^+(E_F) >\},$$

where $j_\alpha^\mu$ denotes the µ-component of the current density operator $\boldsymbol{j}$ for species α with concentration $x_\alpha$ and $G^+(\varepsilon_F)$ is the retarded Green's function at the Fermi energy. Within the KKR-CPA calculations, the local density approximation (LDA)[66] is employed for exchange and correlation. The sensitivity of the results to the exchange-correlation functional is discussed in the Supplementary Information, Section 2, Fig. S1. To study the effect of displacement scattering and spin disorder on $\rho_0$, we used the so-called alloy analogy model (AAM) to perform the configurational average over a discrete set of species-resolved atomic displacements and local moment orientations[31]. To obtain the statistics of the atomic displacements and local moments, we performed standard supercell calculations for a 256-atom special quasi-random structure (SQS)[67] using the projector augmented wave method (PAW)[68] as implemented in the Vienna *ab-initio* simulation package (VASP)[69]. (see the Supplementary Information, Section 1)

**Data availability**

The authors declare that the data supporting this study are available from the corresponding author upon request.

**Acknowledgements** This work is supported by the Energy Dissipation and Defect Evolution (EDDE) center, an Energy Frontier Research Center funded by the U. S. Department of Energy, Office of Science, Basic Energy Sciences under contract number DE-AC05-00OR22725. SM acknowledges fruitful discussions with K. D. Belashchenko, B. C. Sales and K. Jin. SW, SM, and HE would like to thank the DFG (Deutsche Forschungsgemeinschaft) for financial support within the priority program SPP 1538 and the collaborative research centers 689 and 1277.

**Author contributions:** G.M.S conceived this research; S. M. carried out the first-principles calculations and analyzed the results with G.M.S., G.D.S, and S.W.; S.M. and G.M.S wrote the paper, and all authors participated in the discussions and contributed to finalize the paper.

## Additional information

**Supplementary information** accompanies the paper on the *npj* Computational Materials website

**Competing interests**: The authors declare no competing interests.

## References

1. Cantor, B., Chang, I. T. H., Knight, P. & Vincent, A. J. B. Microstructural development in equiatomic multicomponent alloys. *Mater. Sci. Eng. A* **375,** 213–218 (2004).

2. Cantor, B. Stable and metastable multicomponent alloys. in *Annales de chimie* **32,** 245–256 (2007).

3. Miracle, D. B. & Senkov, O. N. A critical review of high entropy alloys and related concepts. *Acta Mater.* **122,** 448–511 (2017).

4. Gludovatz, B. *et al.* A fracture-resistant high-entropy alloy for cryogenic applications. *Science (80-. ).* **345,** 1153–1158 (2014).

5. Zhang, Y. *et al.* Influence of chemical disorder on energy dissipation and defect evolution

in concentrated solid solution alloys. *Nat. Commun.* **6,** 8736 (2015).

6. Diao, H. Y., Feng, R., Dahmen, K. A. & Liaw, P. K. Fundamental deformation behavior in high-entropy alloys: An overview. *Curr. Opin. Solid State Mater. Sci.* **21,** 252–266 (2017).

7. Wu, Z., Bei, H., Otto, F., Pharr, G. M. & George, E. P. Recovery, recrystallization, grain growth and phase stability of a family of FCC-structured multi-component equiatomic solid solution alloys. *Intermetallics* **46,** 131–140 (2014).

8. Jin, K. *et al.* Tailoring the physical properties of Ni-based single-phase equiatomic alloys by modifying the chemical complexity. *Sci. Rep.* **6,** 20159 (2016).

9. Nordheim, L. The electron theory of metals. *Ann. Phys.* **9,** 607 (1931).

10. Linde, J. O. Elektrische Eigenschaften verdünnter Mischkristallegierungen III. Widerstand von Kupfer- und Goldlegierungen. Gesetzmäßigkeiten der Widerstandserhöhungen. *Ann. Phys.* **15,** 219 (1932).

11. Swihart, J. C., Butler, W. H., Stocks, G. M., Nicholson, D. M. & Ward, R. C. First-principles calculation of the residual electrical resistivity of random alloys. *Phys. Rev. Lett.* **57,** 1181 (1986).

12. Ioffe, A. F. & Regel, A. R. Non-crystalline, amorphous and liquid electronic semiconductors. *Prog. Semicond* **4,** 237–291 (1960).

13. Mott, N. F. Conduction in non-crystalline systems-09-the minimum metallic conductivity. *Philos. Mag.* **26,** 1015–1026 (1972).

14. Mooij, J. H. Electrical conduction in concentrated disordered transition metal alloys. *Phys. status solidi* **17,** 521–530 (1973).

15. Gunnarsson, O., Calandra, M. & Han, J. E. Colloquium: Saturation of electrical resistivity. *Rev. Mod. Phys.* **75,** 1085 (2003).

16. Emery, V. J. & Kivelson, S. A. Superconductivity in Bad Metals. *Phys. Rev. Lett.* **74,** 3253–3256 (1995).

17. Hussey, N. E., Takenaka, K. & Takagi, H. Universality of the Mott-Ioffe-Regel limit in metals. *Philos. Mag.* **84,** 2847–2864 (2004).

18. Hartnoll, S. A. Theory of universal incoherent metallic transport. *Nat. Phys.* **11,** 54 (2015).

19. Pierce, F. S., Poon, S. J. & Guo, Q. Electron localization in metallic quasicrystals. *Science* **261,** 737–9 (1993).

20. Sales, B. C. *et al.* Quantum critical behavior in a concentrated ternary solid solution. *Sci. Rep.* **6,** 26179 (2016).

21. Stocks, G. M., Williams, R. W. & Faulkner, J. S. Densities of states in Cu-Rich Ni-Cu alloys by the coherent-potential approximation: Comparisons with rigid-band and virtual-crystal approximation. *Phys. Rev. Lett.* **26,** 253–256 (1971).

22. Mott, N. F. Electrons in transition metals. *Adv. Phys.* **13,** 325–422 (1964).

23. Banhart, J., Vernes, A. & Ebert, H. Spin-orbit interaction and spontaneous galvanomagnetic effects in ferromagnetic alloys. *Solid State Commun.* **98,** 129–132 (1996).

24. Johnson, D. D., Pinski, F. J. & Stocks, G. M. Self-consistent electronic structure of disordered $Fe_{0.65}Ni_{0.35}$. *J. Appl. Phys.* **57,** 3018–3020 (1985).

25. Samolyuk, G. D., Béland, L. K., Stocks, G. M. & Stoller, R. E. Electron--phonon coupling in Ni-based binary alloys with application to displacement cascade modeling. *J. Phys.*

*Condens. Matter* **28,** 175501 (2016).

26. Pettifor, D. G. Theory of energy bands and related properties of 4d transition metals. I. Band parameters and their volume dependence. *J. Phys. F Met. Phys.* **7,** 613 (1977).

27. Staunton, J. B. The electronic structure of magnetic transition metallic materials. *Reports Prog. Phys.* **57,** 1289 (1994).

28. Kübler, J. *Theory of Itinerant Electron Magnetism*. (2000).

29. Okamoto, N. L., Yuge, K., Tanaka, K., Inui, H. & George, E. P. Atomic displacement in the CrMnFeCoNi high-entropy alloy--A scaling factor to predict solid solution strengthening. *AIP Adv.* **6,** 125008 (2016).

30. Van De Walle, A. & Ceder, G. The effect of lattice vibrations on substitutional alloy thermodynamics. *Rev. Mod. Phys.* **74,** 11 (2002).

31. Ebert, H. *et al.* Calculating linear-response functions for finite temperatures on the basis of the alloy analogy model. *Phys. Rev. B* **91,** 165132 (2015).

32. Moriya, T. Ferro-and Antiferromagnetism of Transition Metals and Alloys. *Prog. Theor. Phys.* **33,** 157–183 (1965).

33. Hamada, N. Calculation of a Local-Environment Effect on the Atomic Magnetic Moment in Ferromagnetic Ni--Cu Alloys. *J. Phys. Soc. Japan* **50,** 77–82 (1981).

34. Heine, V. & Samson, J. H. Magnetic, chemical and structural ordering in transition metals. *J. Phys. F Met. Phys.* **13,** 2155 (1983).

35. Dorleijn, J. W. F. & Miedema, A. R. A quantitative investigation of the two current conduction in nickel alloys. *J. Phys. F Met. Phys.* **5,** 487 (1975).

36. Fert, A. & Campbell, I. A. Electrical resistivity of ferromagnetic nickel and iron based alloys. *J. Phys. F Met. Phys.* **6,** 849 (1976).

37. Shi, S., Wysocki, A. L. & Belashchenko, K. D. Magnetism of chromia from first-principles calculations. *Phys. Rev. B - Condens. Matter Mater. Phys.* **79,** (2009).

38. Oguchi, T., Terakura, K. & Hamada, N. Magnetism of iron above the Curie temperature. *J. Phys. F Met. Phys.* **13,** 145 (1983).

39. Gyorffy, B. L., Pindor, A. J., Staunton, J., Stocks, G. M. & Winter, H. A first-principles theory of ferromagnetic phase transitions in metals. *J. Phys. F Met. Phys.* **15,** 1337 (1985).

40. Mu, S. *et al. Importance of magnetic signatures of medium-entropy NiCoMn alloy on thermodynamics and transport (unpublished)*. (2018).

41. Wannier, G. H. Antiferromagnetism. the triangular ising net. *Phys. Rev.* **79,** 357 (1950).

42. Vannimenus, J. & Toulouse, G. Theory of the frustration effect. II. Ising spins on a square lattice. *J. Phys. C Solid State Phys.* **10,** L537 (1977).

43. Hobbs, D., Hafner, J. & Spišák, D. Understanding the complex metallic element Mn. I. Crystalline and noncollinear magnetic structure of $\alpha$-Mn. *Phys. Rev. B* **68,** 14407 (2003).

44. Mohn, P., Schwarz, K., Uhl, M. & Kübler, J. Spin-glass precursors in bcc-manganese. *Solid State Commun.* **102,** 729–733 (1997).

45. Schneeweiss, O. *et al.* Magnetic properties of the CrMnFeCoNi high-entropy alloy. *Phys. Rev. B* **96,** 14437 (2017).

46. Yin, J., Eisenbach, M. & Stocks, G. M. *(unpublished)*. (2018).

47. Jiang, X., Ice, G. E., Sparks, C. J., Robertson, L. & Zschack, P. Local atomic order and

individual pair displacements of Fe 46.5 Ni 53.5 and Fe 22.5 Ni 77.5 from diffuse x-ray scattering studies. *Phys. Rev. B* **54,** 3211 (1996).

48. Lowitzer, S., Koedderitzsch, D., Ebert, H. & Staunton, J. B. Electronic transport in ferromagnetic alloys and the Slater-Pauling curve. *Phys. Rev. B* **79,** 115109 (2009).

49. Lowitzer, S. *et al.* An ab initio investigation of how residual resistivity can decrease when an alloy is deformed. *EPL* **92,** 37009 (2010).

50. Thomas, H. Über Widerstandslegierungen. *Zeitschrift für Phys.* **129,** 219–232 (1951).

51. Jin, K. *et al.* Thermophysical properties of Ni-containing single-phase concentrated solid solution alloys. *Mater. Des.* **117,** 185–192 (2017).

52. Pawel, R. E. & Stansbury, E. E. Specific heat contributions of the ferromagnetic transition in nickel and nickel-copper alloys. *J. Phys. Chem. Solids* **26,** 757–765 (1965).

53. Kroeger, D. M., Koch, C. C., Scarbrough, J. O. & McKamey, C. G. Effects of short-range order on electronic properties of Zr-Ni glasses as seen from low-temperature specific heat. *Phys. Rev. B* **29,** 1199 (1984).

54. Kim, S., Kuk, I. H. & Kim, J. S. Order--disorder reaction in Alloy 600. *Mater. Sci. Eng. A* **279,** 142–148 (2000).

55. Rowlands, D. A. Short-range correlations in disordered systems: Nonlocal coherent-potential approximation. *Reports Prog. Phys.* **72,** 086501 (2009).

56. Minár, J. *et al.* Multiple-scattering formalism for correlated systems: A KKR-DMFT approach. *Phys. Rev. B - Condens. Matter Mater. Phys.* **72,** 045125 (2005).

57. Gludovatz, B. *et al.* Exceptional damage-tolerance of a medium-entropy alloy CrCoNi at

cryogenic temperatures. *Nat. Commun.* **7,** 10602 (2016).

58. Korringa, J. On the calculation of the energy of a Bloch wave in a metal. *Physica* **13,** 392–400 (1947).

59. Kohn, W. & Rostoker, N. Solution of the schrödinger equation in periodic lattices with an application to metallic lithium. *Phys. Rev.* **94,** 1111–1120 (1954).

60. Soven, P. Coherent-potential model of substitutional disordered alloys. *Phys. Rev.* **156,** 809 (1967).

61. The Munich SPR-KKR package. *version 7.7, H. Ebert et al.* http://olymp.cup.uni-muenchen.de/ak/ebert/SPRKKR (2017).

62. Ebert, H., Ködderitzsch, D. & Minár, J. Calculating condensed matter properties using the KKR-Green's function method—recent developments and applications. *Reports Prog. Phys.* **74,** 096501 (2011).

63. Kubo, R. Statistical'Mechanical Theory of Irreversible Processes. I. General Theory and Simple Applications to Magnetic and Conduction Problems. *J. Phys. Soc. Japan* **12,** 570–586 (1957).

64. Greenwood, D. A. The Boltzmann equation in the theory of electrical conduction in metals. *Proc. Phys. Soc.* **71,** 585–596 (1958).

65. Butler, W. H. Theory of electronic transport in random alloys: Korringa-Kohn-Rostoker coherent-potential approximation. *Phys. Rev. B* **31,** 3260 (1985).

66. Perdew, J. P. & Zunger, A. Self-interaction correction to density-functional approximations for many-electron systems. *Phys. Rev. B* **23,** 5048–5079 (1981).

67. Zunger, A., Wei, S.-H., Ferreira, L. G. & Bernard, J. E. Special quasirandom structures. *Phys. Rev. Lett.* **65,** 353 (1990).

68. Blöchl, P. E. Projector augmented-wave method. *Phys. Rev. B* **50,** 17953 (1994).

69. Kresse, G. & Furthmüller, J. Efficiency of ab-initio total energy calculations for metals and semiconductors using a plane-wave basis set. *Comput. Mater. Sci.* **6,** 15–50 (1996).

**Figures**

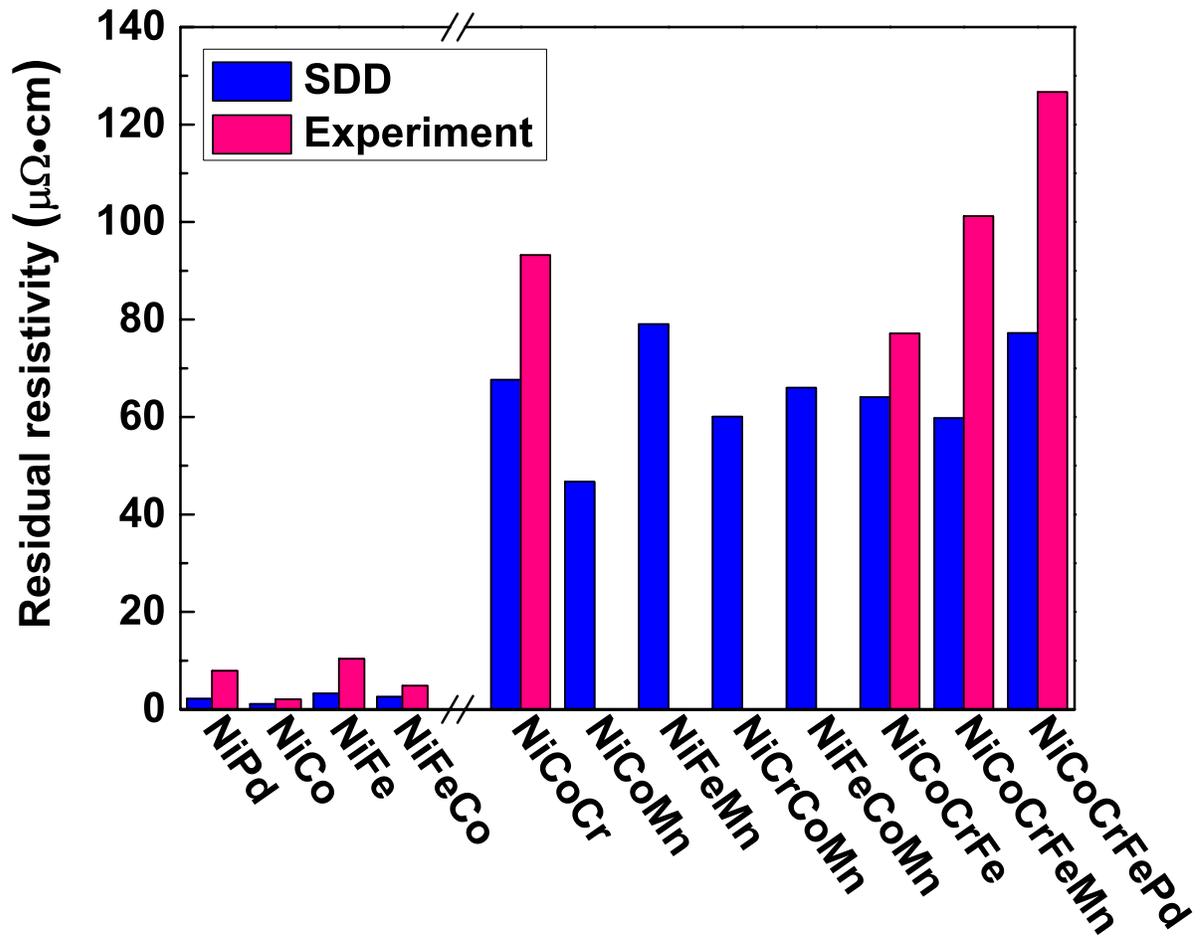

Figure 1 Residual resistivity of Cantor-Wu alloys due to site-diagonal disorder (SDD) (blue bars), compared with experimental data (red bars).

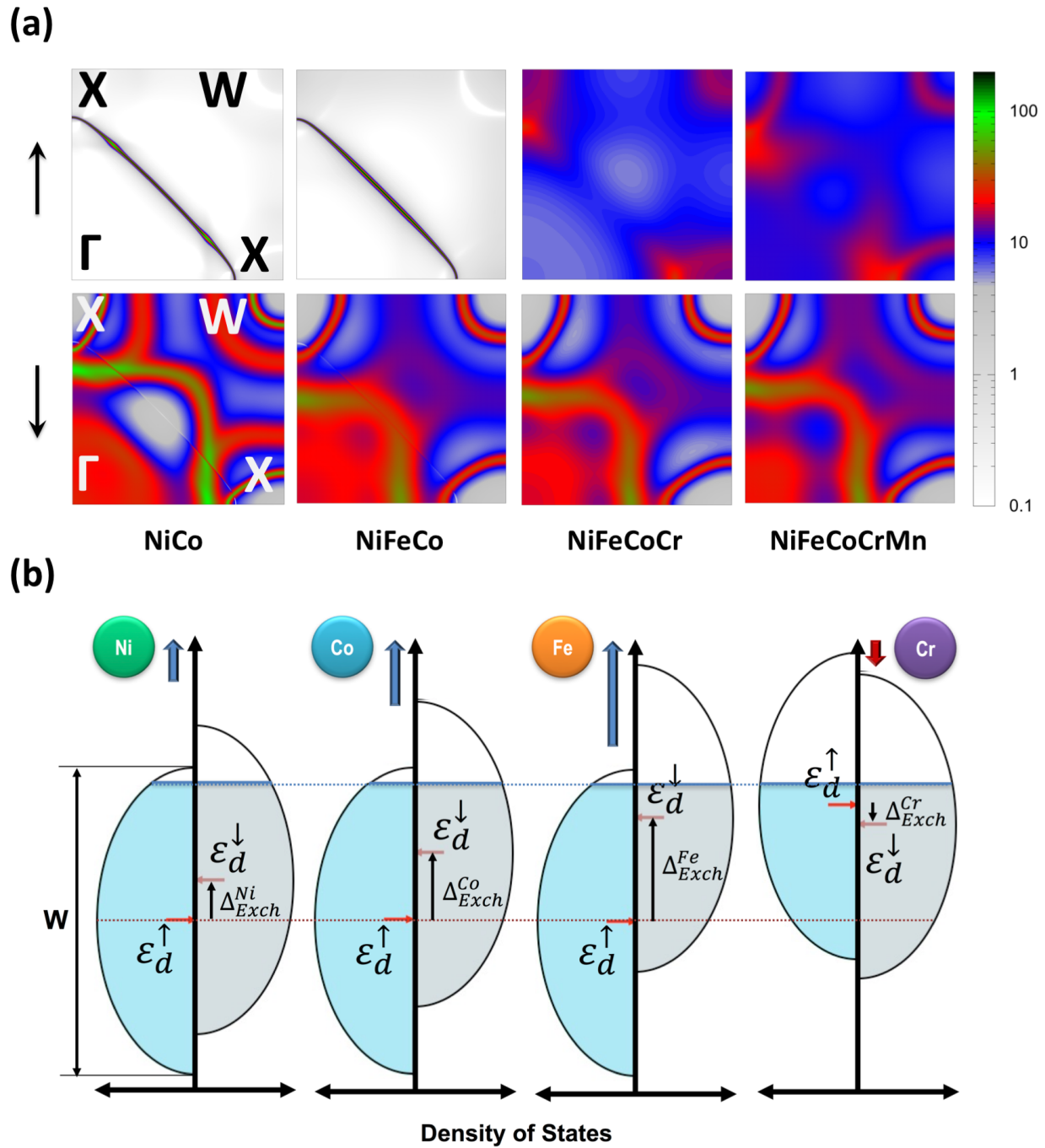

Figure 2 (a) Fermi surface (Γ-X-W plane) in NiCo, NiFeCo, NiFeCoCr, and NiFeCoCrMn alloys (given in arbitrary units). Majority- and Minority- spin channels are distinguished using ↑ and ↓; (b) Magnetic origin of the energy scales for disorder scattering: Simplified depiction of the species-resolved DOS in NiFeCoCr, in which $\varepsilon_d^\sigma$ denotes the *3d* band centers for spin σ(σ = ↑ or ↓). The band center mismatch $\delta_\sigma$ is defined as max{$\varepsilon^\sigma_d(i)$}-min{$\varepsilon^\sigma_d(i)$} with *i* denoting different species. The exchange splitting is denoted as $\Delta_{Exch}:= \varepsilon^\downarrow_d - \varepsilon^\uparrow_d$.

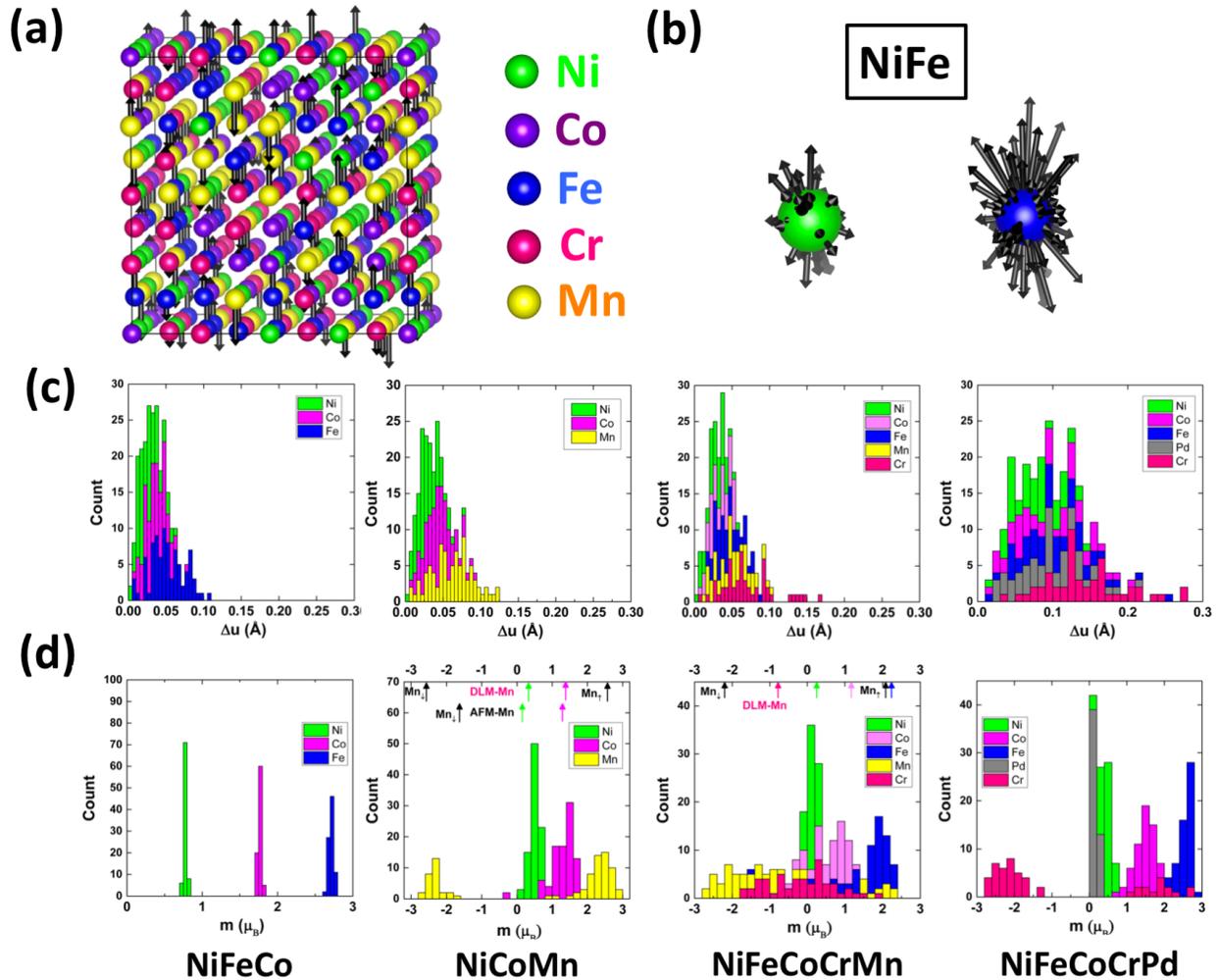

Figure 3 (a) A 256-atom supercell of NiFeCoCrMn with the arrows denoting the local moments. (b) The atomic displacements – with arrows denoting displacement vector away from the ideal positions – on both Ni (green sphere) and Fe (blue sphere) species in a 256-atom supercell of NiFe. Displacement (c) and local moment (d) distributions of each species in NiFeCo, NiCoMn, NiFeCoCrMn, and NiFeCoCrPd, calculated from a 256-atom supercell. The arrows in the moment statistics of NiCoMn and NiFeCoCrMn label the local moment for each species in different states.

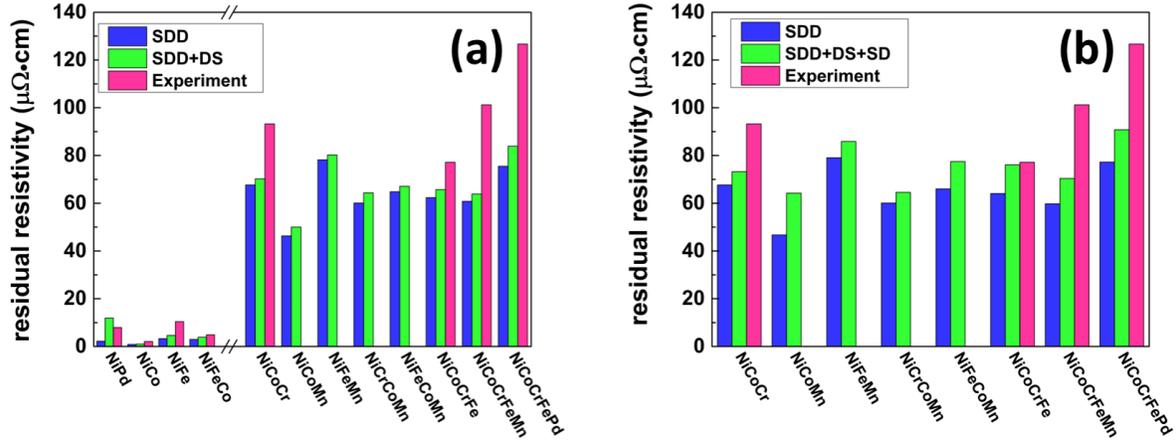

Figure 4 (a) Effect of displacement scattering (DS) on $\rho_0$ in Cantor-Wu alloys; (b) Effect of additional spin disorder (SD) on $\rho_0$ of high-$\rho_0$ alloys.

## Tables:

**Table 1** The total energies ($E_{tot}$, meV/site) and residual resistivity ($\rho_0$, $\mu\Omega\cdot$cm) for the multiple magnetic states in NiCoMn and NiFeCoCrMn from KKR-CPA.

| Alloys | NiCoMn | | NiFeCoCrMn | |
|---|---|---|---|---|
| *States* | $E_{tot}$ | $\rho_0$ | $E_{tot}$ | $\rho_0$ |
| AFM-Mn | 22.5 | 46.7 | 0 | 61.0 |
| DLM | 15.7 | 66.7 | 4.2 | 62.1 |
| DLM-Mn | 0 | 72.0 | 0.12 | 65.7 |